\documentclass[prb, aps,preprint,showkeys,preprintnumbers,superscriptaddress,floatfix]{revtex4-1}

\usepackage{graphicx}		% Include figure files
\usepackage{dcolumn}		% Align table columns on decimal point
\usepackage{bm}				% bold math
\usepackage{amsmath}
\usepackage{amssymb}
\usepackage{amsfonts}
\usepackage{amssymb}
\usepackage{color}
\usepackage{enumitem}

\renewcommand{\deg}{$^\circ$}

\begin{document}

\title{Spin wave emission by spin-orbit torque antennas}

\author{Giacomo Talmelli}
\email[Author to whom correspondence should be addressed. E-mail: ]{\texttt{giacomo.talmelli@imec.be}}
\affiliation{Imec, B-3001 Leuven, Belgium}
\affiliation{KU Leuven, Departement Materiaalkunde, 3001 Leuven, Belgium}
\author{Florin Ciubotaru}
\affiliation{Imec, B-3001 Leuven, Belgium}
\author{Kevin Garello}
\affiliation{Imec, B-3001 Leuven, Belgium}
\author{Xiao Sun}
\affiliation{Imec, B-3001 Leuven, Belgium}
\author{Marc Heyns}
\affiliation{Imec, B-3001 Leuven, Belgium}
\affiliation{KU Leuven, Departement Materiaalkunde, 3001 Leuven, Belgium}
\author{Iuliana P. Radu}
\author{Christoph Adelmann}
\affiliation{Imec, B-3001 Leuven, Belgium}
\author{Thibaut Devolder}
\email[E-mail: ]{\texttt{thibaut.devolder@u-psud.fr}}
\affiliation{Centre de Nanosciences et de Nanotechnologies, CNRS, Universit\'e Paris-Sud, Universit\'e Paris-Saclay, 91405 Orsay cedex, France}

\date{\today}

\begin{abstract}
We study the generation of propagating spin waves in Ta/CoFeB bilayer waveguides by spin-orbit torque antennas and compare them to conventional inductive antennas. The spin-orbit torque was generated by a transverse microwave current across the waveguide. The detected spin wave signals for a transverse in-plane magnetization (Damon-Eshbach configuration) exhibited the expected phase rotation and amplitude decay upon propagation when the current spreading was taken into account. Wavevectors up to about 6\,rad/$\mu$m could be excited by the spin-orbit torque antennas despite the current spreading, presumably due to the non-uniformity of the microwave current. The relative magnitude of generated anti-damping spin-Hall and Oersted fields was calculated within an analytic model and it was found that they contribute approximately equally to the total effective field generated by the spin-orbit torque antenna. The prospects for obtaining a pure spin-orbit torque response are discussed, as are the scaling properties of spin-orbit torque and inductive antennas.
\end{abstract}

\keywords{spin waves, spin-orbit torque, spin pumping, inverse spin Hall effect}

\maketitle

\section{Introduction}

Magnonics is the science of generating, propagating, and detecting spin waves with applications envisioned in wave-based computing \cite{chumak_magnon_2015,Khitun_Wang_2011} or microwave electronics.\cite{MSSW_Review,MSSW_book} Intense research activities in recent years have led to solutions for the routing and the combination of spin waves,\cite{vogt_spin_2012, klingler_design_2014} their amplification, \cite{chumak_magnon_2015, gladii_spin_2016} as well as their detection.\cite{demokritov_brillouin_2001} However, the energy-efficient emission of spin waves remains a difficult task with no consensual solution. For instance, the emission of spin waves by spin-torque oscillators \cite{demidov_direct_2010,Madami_2011, Chen_review} or spin Hall oscillators \cite{collet_generation_2016,Chen_review} lacks the bandwidth and the high spectral purity required for wave-based computing applications, in particular to enable frequency-division multiplexing. The generation of spin waves with a large bandwidth (\textit{i.e.} with a broad range of accessible frequencies/wavelengths) is still mostly achieved using inductive techniques.\cite{liu_spin_2007, wu_fast_2006, vlaminck_spin-wave_2010, gladii_spin_2016} However, these techniques are not energy-efficient due to the large wavelength mismatch between spin waves and electromagnetic waves. Alternative spin wave generation methods using inverse magnetostriction devices \cite{cherepov_electric-field-induced_2014, Barra, Duflou} lack maturity and improved energy efficiency is still to be demonstrated. Overall, an energy-efficient and flexible way to excite spin waves from electrical signals is still lacking.

Spin orbit torques (SOTs)\cite{SOT_Review} generated by electrical currents have recently proven their potential to switch the magnetization in nanomagnets \cite{yu_switching_2014,fukami_spinorbit_2016} on sub-ns timescales. \cite{fukami_sub-ns_2016, baumgartner_time-_2017} In general, SOTs are promising to manipulate the magnetization in nanosystems with large bandwidth. It is thus natural to investigate whether SOTs in the radio frequency (RF) regime can be employed to excite spin waves and potentially replace inductive excitation schemes. 

In this paper, we study the generation of spin waves by RF SOT in thin in-plane magnetized Ta/CoFeB waveguides. The development of a specific device design allowed for the reduction of the parasitic coupling between the spin wave emitter and the collector that results from the direct resistive path between the SOT-providing conductor and the (metallic) magnetic material. Using this approach, electrical spin wave excitation by an SOT-based transducer was experimentally demonstrated and compared to inductive microwave antenna excitation. By adapting a model developed for inductive spin wave excitation and propagation,\cite{ciubotaru_all_2016} the contribution of SOT and Oersted field to the total magnetic field and resulting torque were assessed. Finally, a simple circuit model was derived for SOT antennas to evaluate the scaling behavior and compare it to that of inductive antennas.

\section{Device Concept, Fabrication, and Microwave Behavior}

\begin{figure*}[p]
	\includegraphics[width= 16 cm]{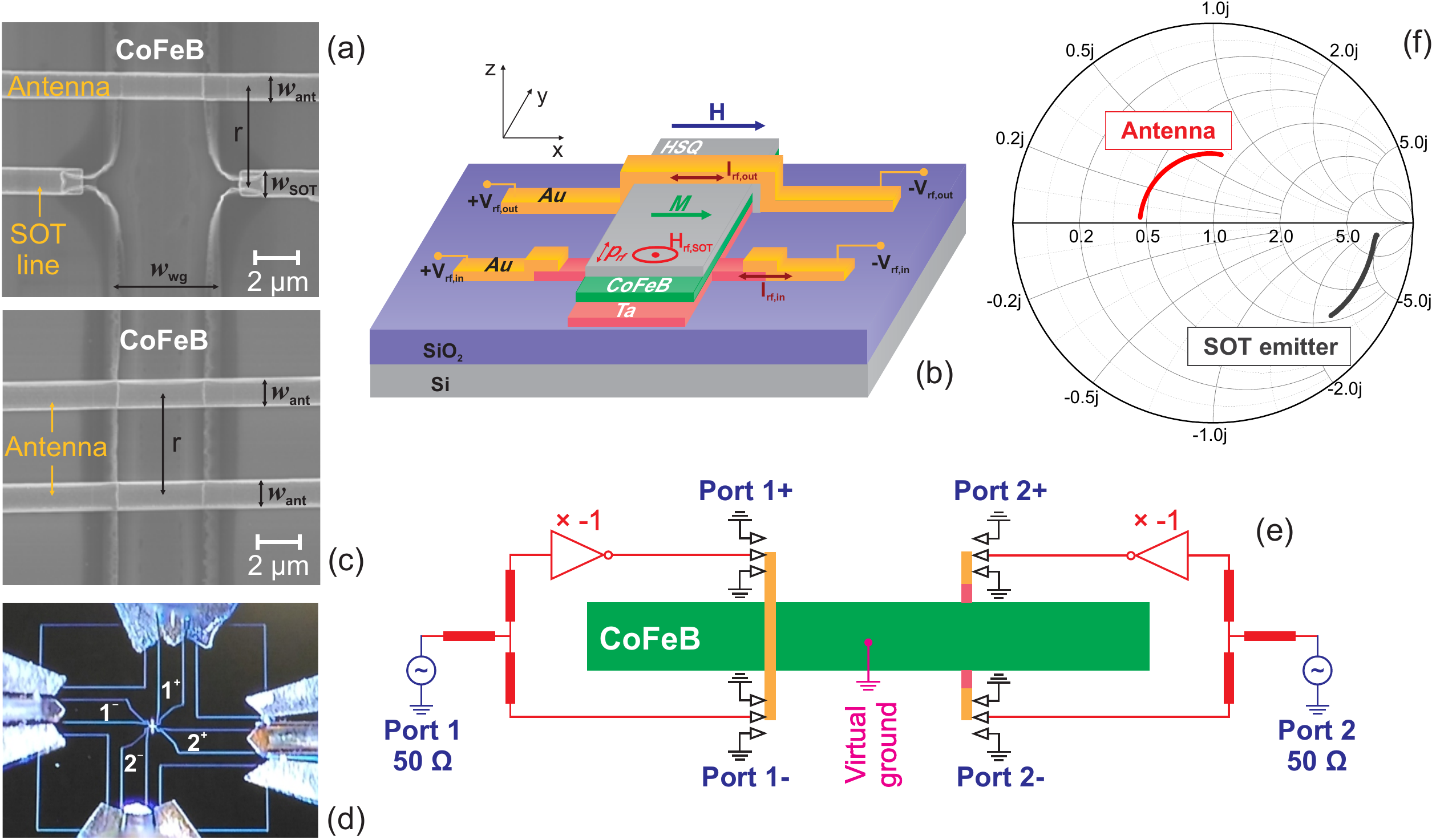}
	\caption{Scanning electron micrographs of devices with (a) an SOT and (c) and inductive antenna as emitter. In both cases, spin waves were detected by an inductive antenna. (b) Sketch of the device illustrating the working principle for the spin wave emission by RF SOTs and detection by an inductive antenna. (d) Optical micrograph of the probing layout with four ports (labeled $1^+$, $1^-$, $2^+$ and $2^-$) connected to four RF probes. (e) Simplified schematic of the virtual-ground configuration used to measure devices including SOT antennas. The red resistances are parts of power dividers. Coupled to the inverters, they form the RF equivalent of balun transformers. (f) Smith chart representation of the device impedance between 0.1 to 20 GHz for devices using inductive (red line) and SOT (black line) antennas. }
	\label{fig:Device_Experiment}
\end{figure*}

Figure~\ref{fig:Device_Experiment} shows the basic structure of the devices and the scheme of the experimental setup. The devices were fabricated by the following processing steps: initially, (in growth sequence, the numbers in brackets indicate film thicknesses in nm) a Ta(8)/CoFeB(5)/MgO(2)/Ta(2) stack was deposited by physical vapor deposition (PVD) in a Canon Anelva system on a 150-nm-thick SiO$_2$ layer grown on 300\,mm Si (100) substrates. After deposition, the stacks were annealed at 300\deg C for 10\,min. A combination of e-beam lithography and ion-beam patterning was employed to define a CoFeB waveguide with a width of $w_\textrm{WG}=5\,\mu$m and SOT antennas with $w_\textrm{SOT}=1\,\mu$m [Fig.~\ref{fig:Device_Experiment}(a)]. E-beam lithography with an hydrogen silsesquioxane (HSQ) hard mask was then used to remove the CoFeB(5)/MgO(2)/Ta(2) layers on top of the SOT antennas outside the waveguide by ion-beam etching (stopping on the Ta(8) layer) to limit shunting by the CoFeB and Ta cap layers. Additional inductive antennas with a width of $w_\mathrm{ant}=1\,\mu$m were defined by e-beam lithography and Ti(10)/Au(100) lift-off on top of the remaining HSQ (about 20\,nm) [Fig.~\ref{fig:Device_Experiment}(a)]. These antennas were used to detect the spin wave signal emitted by the SOT lines. The same lift-off process was also used to contact the SOT lines. A sketch of the devices is depicted in Fig.~\ref{fig:Device_Experiment}(b). In addition, devices with two inductive antennas [Fig.~\ref{fig:Device_Experiment}(c)] were also processed for comparison. The (SOT-)antenna-to-antenna distance $r$ was varied between 1 and 7\,$\mu$m. In all cases, the ends of each (SOT) antenna were connected to coplanar waveguides (CPWs) with a fully symmetric design [Fig.~\ref{fig:Device_Experiment}(d)], allowing each antenna end to float with respect to the electrical ground. Injecting an RF current via the CPWs into the Ta SOT-antenna led to the pumping of an RF spin current into the CoFeB waveguide, generating an out-of-plane RF effective magnetic field (\emph{cf.} Fig.~\ref{fig:Device_Experiment}(e), see the discussion below). 

The typical resistance of an inductive antenna was $R_\mathrm{ant} = 23$\,$\Omega$. By contrast, the substantially thinner SOT antennas were much more resistive, typically $R_\textrm{SOT} = 6$\,$\textrm{k}\Omega$. The microwave behavior of the devices (as expressed by their $S$-parameters) was evaluated using a vector network analyser (VNA). The Smith chart representation of the device impedance [Fig.\ref{fig:Device_Experiment}(f)] indicates that the inductive antennas were essentially resistors ($R_\mathrm{ant} = 23$\,$\Omega$) in series with inductances of 290\,pH at 10\,GHz. The inductances were slightly frequency dependent, decreasing from 320\,pH at 1\,GHz to 195\,pH at 20\,GHz. By contrast, the impedance of an SOT antenna indicated a much higher resistance in parallel with a capacitance. The capacitance was most likely related to parasitic capacitive coupling through the SiO$_2$ layer via the slightly conductive Si substrate . 

The parasitic crosstalk between two inductive antennas [\emph{cf.} Fig.~\ref{fig:Device_Experiment}(c)] was inductive at a level of $-22$\,dB ($|S_{21}| = 0.079$) at 10\,GHz for a distance of $r = 4 $\,$\mu$m. The coupling between the SOT antenna (emitter) and the inductive antenna (receiver) [Fig.~\ref{fig:Device_Experiment}(a)] was capacitive. This could be expected since the SOT Ta wires were connected to the (metallic) Ta/CoFeB waveguides, which in turn were capacitively coupled to the receiving antennas through the dielectric HSQ layer. This led to a capacitive crosstalk of $-24$\,dB at 10\,GHz. As shown in Sec.~\ref{paragraphSOT} below, this parasitic coupling can hide the substantially lower spin wave mediated transmission. To overcome this issue, a virtual ground configuration was used that led to an equipotential CoFeB waveguide. In addition, the configuration also minimized the potential difference between the waveguide and the section of the antenna directly above. The virtual ground configuration was implemented by feeding both ends of the Ta line with a differential voltage, while maintaining both ends of the inductive antenna in a differential situation, thus using four microwave probes and two baluns [Fig.~\ref{fig:Device_Experiment}(e)]. Since the baluns contained power dividers, all data below were corrected for the corresponding losses. This configuration reduced the capacitive crosstalk from $-24$ to $-40$\,dB at 10\,GHz. Optionally, a 35\,dB amplifier (not shown) was used at the detection port to boost the sensitivity. The applied power levels during the experiments were 10\,dBm and $-10$\,dBm for spin wave generation by SOT and inductive antennas, respectively.

\section{Magnetic Film Properties and Spin Hall Angle} 

\label{films}

Before turning to spin wave generation experiments, we discuss the magnetic properties of the CoFeB waveguide as well as the spin-Hall angle measured for the Ta/CoFeB bilayer. Broadband ferromagnetic resonance using a VNA (VNA-FMR) \cite{bilzer_vector_2007} was used to determine the saturation magnetization $M_S$ of the CoFeB film. A value of $\mu_0 M_S = 1.8$\,T was obtained, typical of CoFeB films. The Gilbert damping of the CoFeB layer extracted from VNA-FMR experiments was $\alpha=0.005$. 

The characterization of the SOT in the stack was performed using the second harmonic Hall voltage method.\cite{garello_symmetry_2013, avci_interplay_2014, hayashi_quantitative_2014} For this experiment, separate CoFeB/Ta Hall bar structures were processed by the same lithography approach described above. Following Ref.~\onlinecite{SOT_Review}, contributions by thermal effects were accounted for.\cite{liu_spin-torque_2012, avci_interplay_2014} The effective spin-Hall angle for the entire stack was $\theta_\mathrm{eff} = 0.046 \pm 0.005$. However, part of the current was shunted by the CoFeB waveguide and did not contribute to the spin-Hall effect. Using a parallel resistor model and the measured resistivities of the Ta/CoFeB bilayer (112\,$\mu\Omega$cm) and the Ta layer (160\,$\mu\Omega$cm), $\sim$43\%{} of the current flowed in the Ta layer. After correcting for the shunting effect, the intrinsic spin Hall angle was $\theta = 0.088 \pm 0.009$, which is in line with previous reports.\cite{SOT_Review,fukami_spinorbit_2016,Ta_relative_SOT} Measurements of field-like torques were consistent (within experimental precision) with the Oersted field generated by the RF current using the same above parallel resistor approach. The absence of strong field-like SOTs is consistent with the rather thick films in the stack, in which ``bulk'' effects dominate over interface ones.\cite{FL-SOT}

\section{Reference experiment: all-inductive generation and detection of spin waves}

\begin{figure}[p]
	\includegraphics[width = 10 cm]{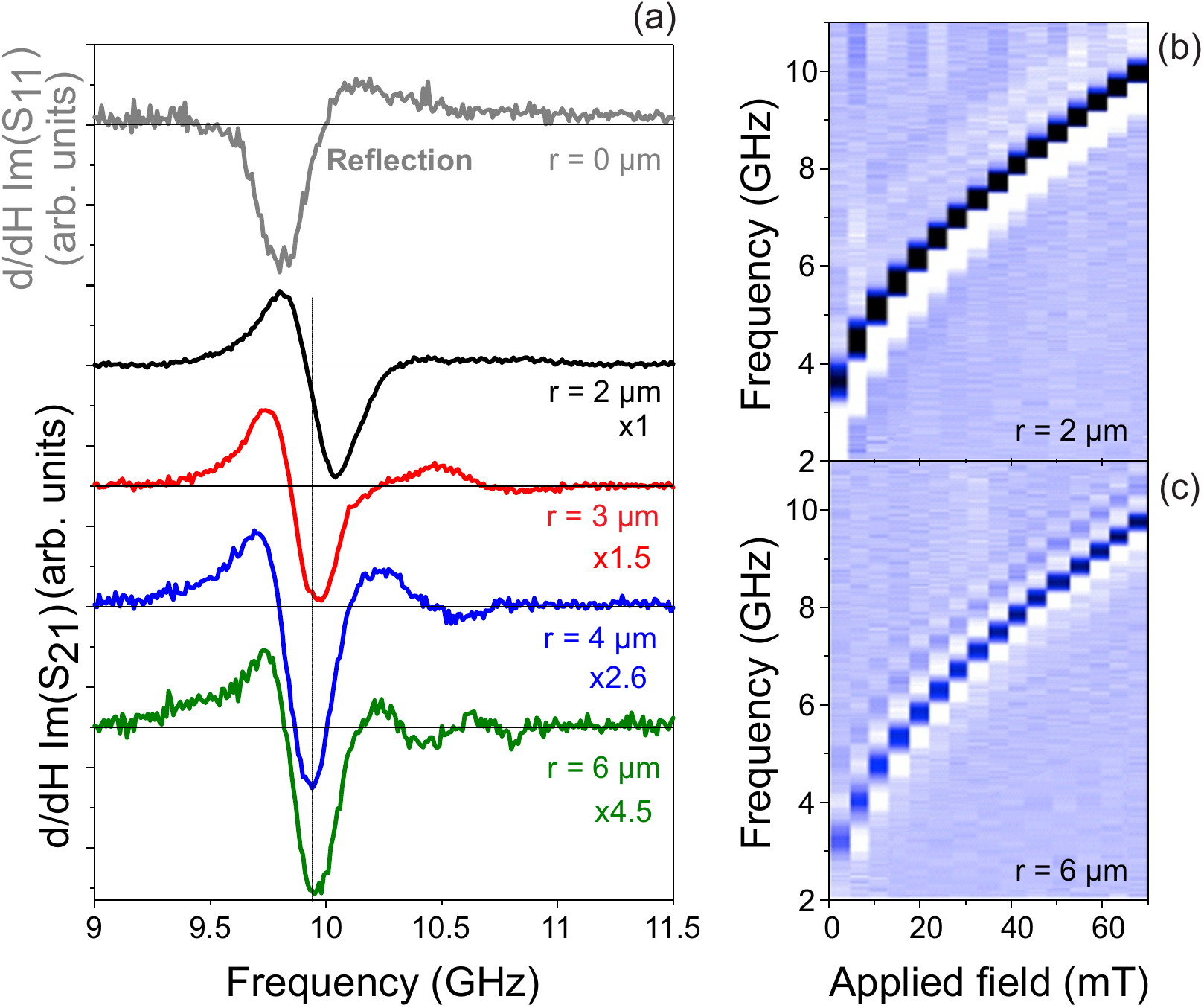}
	\caption{Spin wave signals for devices using all-inductive spin wave emission and detection. The applied field was 70 mT. Top curve: field derivative of the imaginary part of the reflection scattering parameter $S_{11}$. Bottom curves: field derivative of the imaginary part of the transmission scattering parameters $S_{21}$ for propagation distances of 2, 3, 4 and 6\,$\mu$m, respectively. The curve were normalized to each others using the indicated multiplication factors. All data were offset for clarity. Field-frequency signal maps for spin-wave propagation distances of 2\,$\mu$m (b) and 6\,$\mu$ (c), respectively.}
	\label{fig:antenna_emission}
\end{figure}

In a next step, experiments on all-inductive emission and detection of spin waves by conventional antennas were performed to assess the properties of the CoFeB waveguides. The antenna-to-antenna distance $r$ was varied in a range between 2 to 6\,$\mu$m to study spin wave propagation and attenuation. A static magnetic bias field of $\mu_0 H = 70$\,mT was applied transverse to the waveguide (Damon-Eshbach geometry). To assess spin wave related signals, reflection and transmission parameters were defined based on measured scattering parameters $S_{11}$ (reflection) and $S_{21}$ (transmission). To emphasize the spin wave related signals (and to effectively de-embed the parasitics not related to spin waves), magnetic field derivatives of the scattering parameters were used. The reflection parameter $d(\operatorname{Im}S_{11})/dH$ (Fig.~\ref{fig:antenna_emission}, top curve) shows the power absorption due to the excitation of spin waves of all wavevectors compatible with the antenna geometry. The bottom of the band corresponds to the ferromagnetic resonance (FMR) frequency $\omega_\textrm{FMR}/2\pi$. A Kittel fit of the magnetic field dependence of the FMR frequency (not shown) revealed a saturation magnetization of $\mu_0 M_S = 1.8$\,T of the patterned CoFeB waveguide, in keeping with the VNA-FMR results obtained on blanket films (see above). The transmission parameter $d(\operatorname{Im}S_{21})/dH$ (Fig.~\ref{fig:antenna_emission}, lower curves) can be used to deduce the attenuation and the spin wave group velocities. During propagation for a distance $r$ along the waveguide, each spin wave with wavevector $k$ undergoes a phase rotation of $e^{i k r}$ and its amplitude decays by $e^{-r / L_\mathrm{att}}$. This leads to oscillations in the phase-sensitive transmitted signal for larger values of $r$, as shown in Fig.~\ref{fig:antenna_emission}.

The exponential decay of the spin wave during propagation led to a decrease in magnitude of the transmission parameters with $r$. In Fig.~\ref{fig:antenna_emission}, this is visualized by the scaling factors that were needed to compensate for the loss and to obtain constant amplitudes. The dependence of the transmission parameter on the antenna-to-antenna distance $r$ was then used to extract the spin wave attenuation length. It was deduced to be $L_\mathrm{att} = 2.6 \pm 0.3$\,$\mu$m. The Gilbert damping can be expressed by $\alpha=(\gamma_0 M_S t_\mathrm{FM}) / (2 \omega_\textrm{FMR} L_\mathrm{att})$ with $\gamma_0$ the gyromagnetic ratio and $t_\mathrm{FM}$ the thickness of the ferromagnetic CoFeB waveguide. From the measured $L_\mathrm{att}$ and $t_\mathrm{FM} = 5$\,nm, a value of $\alpha = 0.005 \pm 0.0005$ could be extracted, in excellent agreement the VNA-FMR results obtained on blanket films (see above). This indicated that the damping in the CoFeB waveguide was not affected by the device patterning processes. The observation of spin waves in a 5-nm-thin layer after a propagation distance of 6\,$\mu$m is thus a direct consequence of the low damping in the CoFeB film that was maintained after device fabrication.

\section{Excitation of spin waves by radio-frequency spin-orbit torques}

\label{paragraphSOT}

Subsequently, the generation of spin waves by SOT antennas was studied. In all experiments below, spin wave detection was achieved by inductive antennas at a distance of $r = 4$\,$\mu$m for better comparison with the all-inductive spin wave generation and propagation studies above. 

\subsection{Phase rotation and propagation distance of spin waves emitted by SOT antennas\label{SubSec_SOT_SW}}

\begin{figure}[p]
	\includegraphics[width=8 cm]{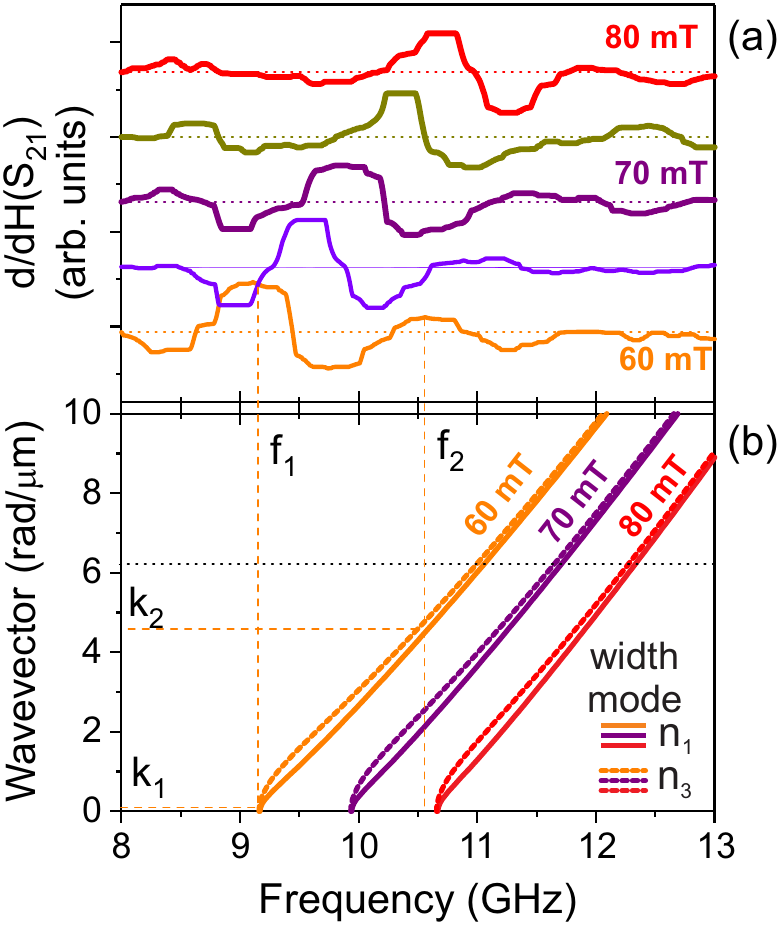}
	\caption {(a) Spin wave transmission signals due to spin wave generation by an SOT antenna and detection by an inductive antenna at a nominal distance $r = 4 ~\mu\textrm{m}$ for several applied transverse magnetic bias fields. All data were offset for clarity. (b) Spin wave dispersion relations of the fundamental and the third-order width modse\cite{kalinikos_theory_1986} calculated for the material parameters specified in the text and a waveguide width of $w_\mathrm{WG} = 5$\,$\mu$m. The horizontal black dashed line indicates the maximum wavevector that can be detected by an antenna of effective width of 1\,$\mu\textrm{m}$.  The thin yellow dashed lines indicate the approximate wavevectors and the approximate frequencies of spin waves responsible for two successive extrema of the transmission coefficient for a bias field of 60 mT.}
	\label{fig:SOT_emission}
\end{figure}

Figure \ref{fig:SOT_emission}(a) shows transmission parameters for spin wave generation by an SOT antenna. The transmission parameter $d(\operatorname{Im}S_{21})/dH$ was defined in the same way than for the all-inductive experiments above. Different static magnetic bias fields were applied transverse to the CoFeB waveguide (Damon-Eshbach geometry, again identical to the all-inductive experiments above), as specified in Fig.~\ref{fig:SOT_emission}(a). A spin wave related transmission signal was clearly visible for every applied field. As in the all-inductive experiments above, the FMR frequency (the bottom of the band) increased  with increasing magnetic bias field in agreement with the material parameters. Above the FMR frequency, spin waves with nonzero wavevector $k$ were excited and underwent phase rotation during propagation, leading to the oscillation of the transmission parameter $d(\operatorname{Im}S_{21})/dH$ [Fig.~\ref{fig:SOT_emission}(a)]. 

However, a closer look at the signal and the comparison with all-inductive signals (Fig.~\ref{fig:SOTversusINDUCTIVE}) indicate significant differences in the behavior. In particular, the period of the oscillations of the transmission signal was significantly smaller for the SOT antenna than for the inductive antenna. The period of the oscillations is determined by $kr'$ with $k$ the wavevector of the propagating spin waves and $r'$ the propagation distance between the source and the detector. The different period thus means that the SOT antenna either excited spin waves with a different dispersion relation or that the effective propagation distance was much smaller than for inductive excitation despite the identical nominal center-to-center distance between source and detection antennas.

\begin{figure*}[p]
\begin{center}
	\includegraphics[width=16 cm]{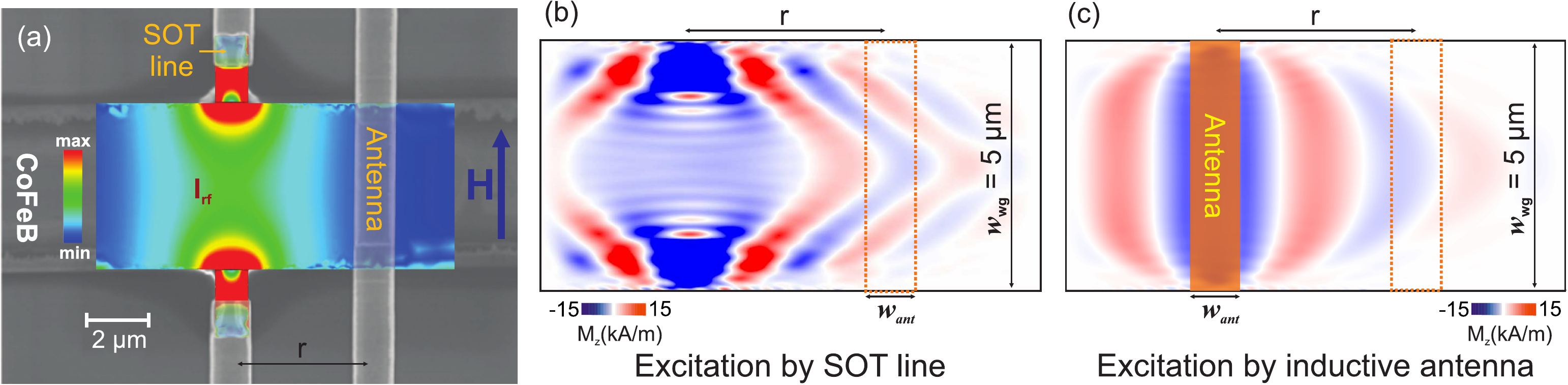}
	\end{center}
	\caption {(a) Calculated distribution of the RF current in an SOT antenna ($f = 11.2$\,GHz) indicating highly non-uniform current spreading. Micromagnetic simulations of spin wave excitation at 11.2\,GHz (b) by an SOT antenna with the current distribution shown in (a) and, for comparison, (c) by an inductive antenna. The images show the $M_z$-component after excitation for 5.25\,ns. The applied magnetic bias field was 80\,mT.}
	\label{fig:calculations}
\end{figure*}

To shed light on this issue, we have calculated the RF current distribution in the SOT antenna using three-dimensional full-wave electromagnetic simulations (Ansys HFSS 18.1). The result is shown in Fig.~\ref{fig:calculations}(a) and indicated (not unexpectedly) significant current spreading in the CoFeB/Ta waveguide. In addition, current concentration occurred near the connections of the leads and the waveguide. Thus, the current distribution in the SOT antenna was highly non-uniform. In a next step, the spin wave emission by the SOT-line was studied by micromagnetic simulations\cite{OOMMF} using a similar distribution of the excitation fields as shown in Fig.~\ref{fig:calculations}(a). As discussed in the next section, the effective magnetic field of the SOT antenna was directed along $\bm{e}_y + \bm{e}_z$, with $\bm{e}_y$ and $\bm{e}_z$ the unit vectors along the $y$- and $z$-directions, respectively. The simulated area of $10\,\mu\mathrm{m} \times 5\,\mu\mathrm{m} \times 5\,\mathrm{nm}$ was discretized using a uniform mesh with a cell size of $5 \times 5 \times5$\,nm$^3$. The experimentally determined magnetic parameters, such as the saturation magnetization, the bias field, and the Gilbert damping, were used ($\mu_0 M_S = 1.8$\,T, $\mu_0 H = 80$\,mT, $\alpha = 0.005$), while a typical value for the CoFeB exchange stiffness constant was assumed ($A = 18.5$\,pJ/m). Furthermore, a gradually increasing Gilbert damping towards the both ends of the simulated structure was considered to avoid the spin wave reflection. For comparison, simulations were also performed for spin wave excitation by an inductive antenna with a uniform current distribution and a resulting Oersted field along $\bm{e}_y$. 

Figure \ref{fig:calculations}(b) shows the spatial distribution of the $M_z$ component after excitation by an SOT antenna for 5.25\,ns at a frequency of 11.2\,GHz. The resulting mode pattern was consistent with a third-order width mode that was preferentially excited due to the non-uniformity of the driving RF magnetic field with higher torques at the edges of the waveguide due to higher local currents. By contrast, the excitation by the rather uniform Oersted field generated by an inductive antenna led to the excitation of the fundamental width mode, as shown in Fig.~\ref{fig:calculations}(c). We note that the excitation of the third-order width mode (with respect to the fundamental mode) can be expected to lead to a reduction of the detection efficiency by an inductive antenna [\emph{cf.} the red boxes in Figs. \ref{fig:calculations}(b) and (c)]. 

We now turn to the quantitative assessment of $kr'$ and the apparent effective propagation distance $r'$. Figure~\ref{fig:SOT_emission}(b) shows calculated dispersion relations \cite{kalinikos_theory_1986} for different magnetic bias fields between 60 and 80\,mT and for both the fundamental and third-order width modes. The relative phase shift as a function of excitation frequency can be obtained by determining the frequencies at which the oscillations have extrema or zero-crossings. For example, a relative phase shift of $2\pi$ occurs for frequencies at two successive maxima of the transmission parameter $d(\operatorname{Im}S_{21})/dH$. The corresponding wavevectors can be obtained from the appropriate dispersion relations of the third-order and fundamental width modes for SOT and inductive antenna excitation, respectively. A linear regression of the phase shift as a function of wavevector led to an effective propagation distance for SOT excitation of $r' = 1.3\pm 0.2$\,$\mu$m, much smaller than the nominal center-to-center distance between the SOT antenna and the receiving inductive antenna of 4\,$\mu$m. We note that the analysis using dispersion relations of the fundamental width mode would lead to nearly identical results (within experimental error), indicating that the difference in spin wave dispersion for excitation by SOT or inductive antennas cannot account for the much smaller propagation distance. By contrast, an analogous analysis of the oscillation period for all-inductive spin wave transmission leads to $r' = 2.8\pm 0.2$\,$\mu$m, rather close to the nominal (edge-to-edge) antenna distance. A slightly smaller value might be due to the extension of the Oersted field around the inductive antenna that leads to an effective source width that is slightly larger than $w_\mathrm{ant}$.

The much larger deviation from the geometric emitter-detector distance for SOT antennas can be attributed to current spreading in the (conductive) area underneath the Ta/CoFeB waveguide between the SOT antenna connections [\emph{cf.} Fig.~\ref{fig:calculations}(a)] since the Ta layer extends underneath the entire CoFeB waveguide. Additional micromagnetic simulations (not shown) indicate approximately in-phase behavior (\emph{i.e.} a relative phase difference of $2\pi$) at the position of the receiving antenna for excitation at 11.2 and 10.9\,GHz. Using the same analysis than for the experimental data, this corresponds to an effective propagation distance of $r' \approx 1$\,$\mu$m. This value is in reasonable quantitative agreement with the experimental results and clearly indicates the the different period of the oscillations for excitation by SOT or inductive antennas can be ascribed to a shortened propagation distance due to current spreading in the SOT antenna.

Using the dispersion relations in Fig.~\ref{fig:antenna_emission}(b), the data indicate spin wave excitation by SOT antennas up to wavevectors of about 6\,rad/$\mu$m, \emph{i.e.} wavelengths of the order of 1\,$\mu$m. This is close to the limit for detection by a 1-$\mu$m-wide antenna. The relatively large bandwidth desipte the large effective emitter area due to current spreading may be attributed to the strong non-uniformity of the current distribution. However, because to the rather small signal-to-noise ratio for SOT antennas because of the power dividers present in the measurement chain [Fig.~\ref{fig:Device_Experiment}(e)], the modulation of the transmission coefficient due to the excitation of spin waves by SOT stayed near the noise of the VNA. This renders impossible any further quantitative analysis of the shape of the transmission signal for SOT antennas. Thus, the quantitative modeling of detailed line shapes of transmission signals using SOT antennas and the fully quantitative determination of the bandwidth, along the lines of previous work on all-inductive spin wave spectroscopy,\cite{ciubotaru_all_2016} are beyond the scope of this paper. However, below, we will assess the different contributions to the total torque exerted by a current in an SOT-antenna, and estimate their relative importance. 

\subsection{Torques in SOT antennas}

An RF current $I_\mathrm{RF}$ in a SOT-antenna of width $w_\textrm{SOT}$ in contact with a magnetic waveguide with thickness $t_\mathrm{FM}$ generates both SOTs as well as torques due to the associated RF Oersted field. The quantitative understanding of the spin wave excitation thus requires the determination of the relative strength of these contributions. In the following, we use a simple analytic model to compare the magnitudes of the Oersted field and the SOT as well as their relative excitation efficiencies. Assuming a uniform current density located only in the Ta antenna, the in-plane component of the local Oersted field $\bm{H}_\mathrm{Oe}$ can be described approximately by

\begin{equation}
\bm{H}_\mathrm{Oe} \approx \frac{1}{2 w_\textrm{SOT} } I_\mathrm{RF} ~\bm{e}_y,
\label{HOe_SOT}
\end{equation}

\noindent with $\bm{e}_y$ the unit vector along the $y$-direction, \emph{i.e.} along the waveguide [\emph{cf.} Fig.~\ref{fig:Device_Experiment}(b)]. The dynamics of the normalized magnetization $\bm{m}$ can then be described by the Landau-Lifshitz-Gilbert (LLG) equation including the torque due to the spin-Hall effect, 

\begin{equation}
\frac{d\bm{m}}{dt} = \gamma_0 \bm{H}_\mathrm{Oe}\times \bm{m} + \alpha \bm{m}\times \frac{d\bm{m}}{dt} + \gamma_0 c (\bm{m} \times \bm{p})\times \bm{m},
\end{equation}

\noindent with $c = \frac{\hbar \theta}{2 e \mu_0 M_S t_\mathrm{FM}}J_\mathrm{RF}$ the SOT efficiency, $\theta$ the spin-Hall angle, and $\bm{p}$ representing the spin polarization orientation of the injected spin current. Here, we neglect field-like SOTs as they have been found to be negligible in our structures (see Sec.~\ref{films}). The SOT term in the LLG equation is thus proportional to $(\bm{m} \times \bm{p})\times \bm{m}$ and is referred to as the anti-damping term. Using a small signal approximation for spin waves $\bm{m} = \bm{m_0} + \bm{\Delta m}$, the anti-damping term becomes to first order $\left[\left( \bm{m_0} + \bm{\Delta m}\right)\times\bm{p}\right]\times\left( \bm{m_0} + \bm{\Delta m}\right) \approx (\bm{m_0} \times \bm{p})\times \bm{m_0}$. In our geometry, $\bm{m_0}\times \bm{p} = \bm{e}_z$ and therefore, the anti-damping SOT can be considered as the torque due to an effective field 

\begin{equation} 
\bm{H}_\mathrm{{SOT}} \approx \frac{t_{\textrm{SHE}}}{t_\mathrm{SOT}} \frac{1}{2 w_\textrm{SOT} } I_\mathrm{RF} ~\bm{e}_z,
\label{H_SOT}
\end{equation}

\noindent with $\bm{e}_z$ the out-of-plane unit vector, $t_\mathrm{SOT}$ the thickness of the SOT antenna, and $t_{\textrm{SHE}}$ a characteristic length scale given by 

\begin{equation}
t_\textrm{SHE} =  \frac{\hbar}{e}  \frac{\theta}{\mu_0 M_S t_\mathrm{FM}}.
\label{CharactersiticThickness}
\end{equation}

\noindent $t_\textrm{SHE}$ has no geometrical meaning but describes the magnitude of the spin-orbit torque. In our specific case, considering $\theta = 0.088$ (as only the current in the Ta antenna is considered), $t_\textrm{SHE} = 6.6$\,nm. 

The highly similar forms of Eqs.~(\ref{HOe_SOT}) and (\ref{H_SOT}) allow for a straightforward comparison of the magnitude (per current) of the (effective) magnetic fields due to the spin-Hall effect and the Oersted field. Considering the same parallel resistor model as in Sec.~\ref{films} for the SOT antenna and the waveguide, the RF current generates both normalized Oersted (with strength $h_\textrm{Oe}$) and effective anti-damping SOT (with strength $h_\textrm{SOT}$) magnetic fields with a ratio of $h_\textrm{SOT}/h_\textrm{Oe} = t_\mathrm{SHE}/t_\mathrm{SOT}$. In our devices, $h_\textrm{SOT}/h_\textrm{Oe} \sim 1$ and thus, the Oersted field and the anti-damping SOT effective field contributed approximately equally. 

However, the effective fields have different orientations and thus excite magnetization dynamics according to different components of the susceptibility tensor $\bm{\chi}$. In the presence of an RF magnetic field with components $H_y$ and $H_z$, the dynamic response of the magnetization at the FMR frequency can be written as

\begin{equation}
\left( \begin{array}{cc}  m_y \\  m_z \\ \end{array} \right) 
=
-\frac{i}{\alpha  (2 {H_x}+{M_s})} 
\left( \begin{array}{cc}   k & i  \\  i  & \frac{{1}}{k} \\ \end{array} \right)
\left( \begin{array}{cc}  H_y\\  H_z \\ \end{array} \right),
\label{SOT_vs_Oe_comparison}
\end{equation}

\noindent with $k = \sqrt{(H_x + M_S)/H_x}$ the ellipticity of the precession of the magnetization at the FMR frequency. Hence, the complex amplitude of the dynamic response scales with $i \left| H_\textrm{SOT}\right| + {  k |H_\textrm{Oe}|} $ at the FMR frequency. In our experiments, the ellipticity was $k \sim 5$. As a result, the torque due to the Oersted field (along $\bm{e}_y$) was enhanced by the ellipticity with respect to the anti-damping SOT (effective field along $\bm{e}_z$) and dominated the dynamic magnetic response, although the effect of the SOT was not negligible.

\begin{figure}[p]
\begin{center}
	\includegraphics[width=8 cm]{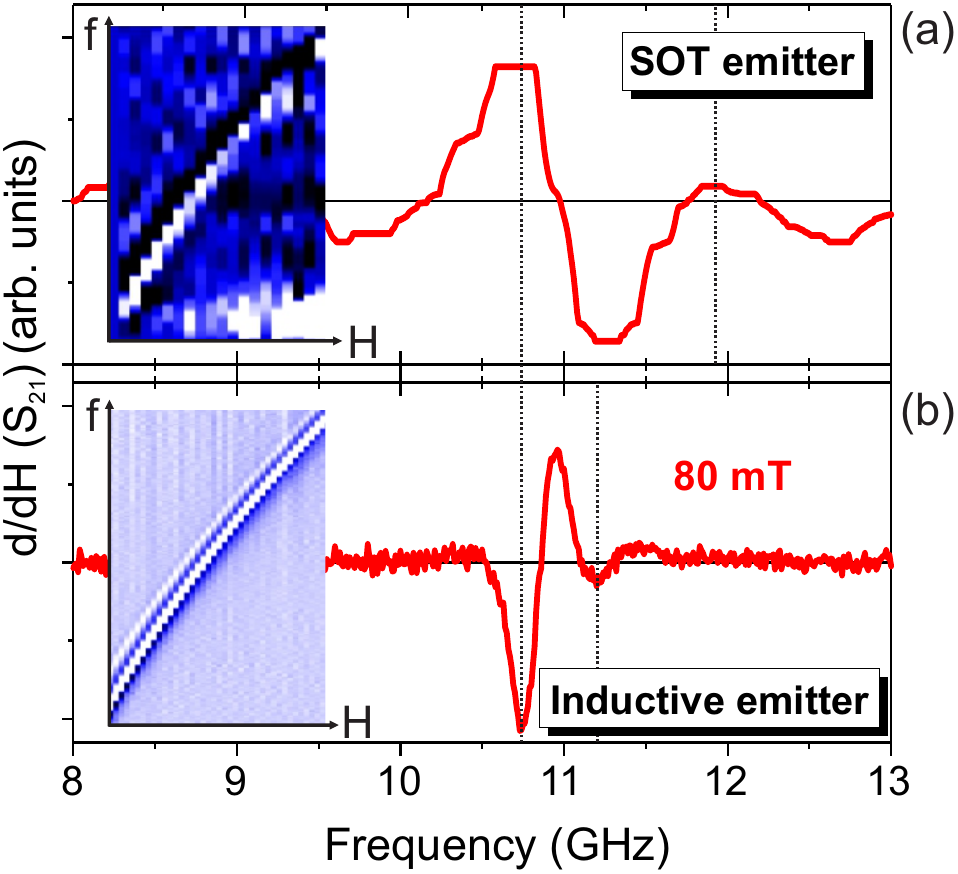}
	\end{center}
	\caption {Compared efficiencies of the spin wave generation by spin-orbit torque antennas [panel (a)] and inductive antennas [panel (b)]. The data are the field derivative of the forward transmission scattering parameter collected by an inductive antenna for emitter-to-detector center-to-center distance of 4 $\mu\textrm{m}$ and an applied field of 80 mT. Inset: field-frequency signal maps in the range of 52-145\,mT and 8-15\,GHz corresponding to the spin waves emitted by the two transducers. }
	\label{fig:SOTversusINDUCTIVE}
\end{figure}

Equation~(\ref{SOT_vs_Oe_comparison}) has also implications for the phase of the magnetization response. Effective fields along $\bm{e}_z$ (\emph{e.g.} the anti-damping SOT field) generate spin waves with a global phase difference of $\pi/2$ with respect to field-like (Oersted) torques with the driving field along $\bm{e}_y$. A quantitative analysis of the phase of the experimental data in Fig.~\ref{fig:SOTversusINDUCTIVE}(a) is difficult due to the low signal-to-noise ratio. The extrapolation of the linear regression of relative phase shift \emph{vs.} wavevector was consistent with the expected (rather small) phase shift due to the mixed SOT and Oersted field excitation. However, the experimental error was too large for a final conclusion and samples with larger SOT contributions will be required to experimentally confirm this point. Note that the observed opposite sign (a global phase shift of $\pi$) of the responses due to the SOT and the Oersted field (\emph{cf.} Fig.~\ref{fig:SOTversusINDUCTIVE}) was due to the fact that the SOT antenna was below the CoFeB waveguide while the inductive antenna was above. 

Equations~(\ref{CharactersiticThickness}) and (\ref{SOT_vs_Oe_comparison}) indicate the conditions, under which a pure SOT response can be obtained and the Oersted contribution becomes negligible. Thinner magnetic waveguides with lower saturation magnetization will both increase $t_\mathrm{SHE}$. In addition, much lager spin-Hall angles can be obtained using $\beta$-W \cite{beta-W} or topological insulators \cite{SOT_TI, SOT_TI2} to generate SOTs. Thus, values of $h_\textrm{SOT}/h_\textrm{Oe} \geq 10$ are clearly within reach by optimizing the structure of the SOT antenna.

\section{Scaling potential of SOT and inductive antennas}

The above discussion allowed to compare quantitatively the effects of Oersted fields and SOTs in an SOT antenna. In the following, we will discuss the transduction efficiency and the scaling potential of SOT antennas and compare it to conventional inductive antennas. This requires the assessment of the power transfer from a source with output impedance $R_S$ ($= 50$\,$\Omega$ in our case). The current in a load antenna with impedance $R_L + iX_L$ can then be written as\cite{RFBOOK}  

\begin{equation}
I_{RF} = \frac{\sqrt{2 R_S \times P_{0}}}{\sqrt{\left(R_S + R_L\right)^2 + X_L^2}}
\label{I_rf}
\end{equation}

\noindent Here, $P_0$ denotes the power delivered to a load with impedance $R_0$, \emph{i.e.} matched to the output impedance of the source. 

We now define a figure of merit (FOM) for spin waves transducers as $\left|\bm{\chi} \bm{H}\right|^2/P_0$, \emph{i.e.} the ratio of the spin wave intensity and the source power. In an ideal SOT antenna, the Oersted contribution is negligible and the spin wave generation is dominated by the anti-damping SOT. Furthermore, for a typical SOT antenna, $R_L \gg X_L$ and $R_L \gg R_S$. We also neglect the radiation resistance due to spin wave excitation (weak coupling), which is equivalent to the assumption that dissipation in the system occurs mainly by Ohmic heating and not by spin wave emission and decay. A lumped model for the Ohmic load impedance together with Eq.~(\ref{H_SOT}) leads then to

\begin{equation}
\mathrm{FOM}_\mathrm{SOT} \propto \left( \frac{1}{\alpha k^2} \frac{t_{\textrm{SHE}}}{\rho\ell}\right)^2,
\label{FOM_SOT}
\end{equation}

\noindent with $\rho$ the resistivity of the SOT antenna and $\ell$ its length (typically close to the waveguide width). This shows that the FOM can be increased strongly by geometrical scaling of the spin wave devices, \textit{i.e} by reducing $\ell$. As discussed above, increasing $t_{\textrm{SHE}}$ (\emph{e.g.} by increasing the spin-Hall angle $\theta$ or by decreasing the magnetic film thickness $t_\mathrm{FM}$) will also lead to an increase of the FOM, as will the reduction of the damping $\alpha$ and the ellipticity $k$. It is worth noting that reducing $t_\mathrm{FM}$ will also lead to a non-negligible contribution of field-like SOTs due to interface effects,\cite{FL-SOT} which scales with $1/k$ (instead of $1/k^2$) and will thus further increase the FOM of SOT antennas. Additional possibilities for improvements are a lower resistivity $\rho$ of the antenna, as well as a lower saturation magnetization $M_S$, which will both increase $t_{\textrm{SHE}}$ and decrease $k$. 

It is instructive to compare this result with the FOM for an inductive antenna. The same approach (with the same approximations) finds that the inductive antenna FOM scales with the antenna thickness $t_\mathrm{ant}$ instead of $t_{\textrm{SHE}}$. Therefore, a geometric scaling (\emph{i.e.} decreasing $t_\mathrm{ant}$ and $\ell$ by the same factor) will not lead to an improvement of the FOM. Hence, for nanoscale spin waves transducers, where the thickness of the antenna might be strongly limited, SOT antennas might ultimately outperform conventional inductive antennas.

\section{Conclusion}

In conclusion, we have used all-electric broadband microwave measurements to study the generation of spin waves in a thin in-plane magnetized Ta/CoFeB waveguide by a transverse current in an SOT antenna. For the stack, a spin-Hall angle of $\theta = 0.088$ was found when only the current in the Ta layer was considered. For spin waves experiments, a specific virtual-ground measurement setup was developed to reduce the parasitic coupling between the SOT antenna (the spin wave emitter) and the inductive spin wave detector that resulted from the direct conductive path between emitter and detector. The setup relied on two baluns that allowed to keep the Ta/CoFeB waveguide at a constant potential, thereby suppressing the capacitive crosstalk. Reference experiments were conducted using all-inductive excitation and detection of spin waves and a conventional measurement setup.

For excitation by both SOT and inductive antennas, the detected spin wave signals exhibited the expected non-reciprocity, phase rotation, and amplitude decay upon propagation. For SOT antennas, the calculated spreading of the current in the Ta/CoFeB waveguide needed to be considered, leading to a significantly shorter effective propagation distance than for inductive antennas with the same nominal antenna-to-antenna distance. This was confirmed by micromagnetic simulations, which also allowed to elucidate the effects of the current non-uniformity on the spin wave mode.

An analytic model was developed to distinguish between effects due to the anti-damping spin-Hall effective field and the Oersted field both generated by the current in the Ta layer. The model found that the magnitude of the spin-Hall and Oersted fields was approximately equal in our samples. However, taking the susceptibility into account, which differs for spin wave excitation by SOTs and Oersted fields due to their different orientations, the effect of the Oersted field was dominant. The analytic model allowed to predict that a dominant SOT response can be obtained by maximizing the spin-Hall angle as well as minimizing the thickness and saturation magnetization of the ferromagnetic waveguide within experimentally accessible ranges. Finally, a circuit model for SOT and inductive antennas was developed, which allowed for the assessment of their scaling potentials. A better scalability of SOT antennas was found, which is related to the fact that the effective spin-Hall field is proportional to the \emph{current density} whereas the Oersted field is proportional to the \emph{current}. Therefore, SOT antennas might outperform inductive antennas at small dimensions.

\begin{acknowledgments}

This work has been supported by imec's industrial affiliate program on beyond-CMOS logic and by a public grant overseen by the French National Research Agency (ANR) as part of the ``Investissements d'Avenir'' program (Labex NanoSaclay, ANR-10-LABX-0035). The authors would also like to thank imec's clean room technical support and especially Nicol\`{o} Pinna (imec) for the e-beam lithography.

\end{acknowledgments}

\clearpage

\clearpage

\clearpage

\clearpage

\begin{thebibliography}{99}

\bibitem{chumak_magnon_2015} A.V. Chumak, V.I. Vasyuchka, A.A. Serga, and B. Hillebrands, Nature Phys. \textbf{11}, 453 (2015).

\bibitem{Khitun_Wang_2011} A. Khitun and K.L. Wang, J. Appl. Phys. \textbf{110}, 034306 (2011).

\bibitem{MSSW_Review} W.S. Ishak, Proc. IEEE \textbf{76}, 171 (1988).

\bibitem{MSSW_book} P. Kabo\v{s} and V.S. Stalmachov, \textit{Magnetostatic Waves and Their Application} (Springer, Dordrecht, 1994).


\bibitem{vogt_spin_2012} K. Vogt, H. Schultheiss, S. Jain, J.E. Pearson, A. Hoffmann, S.D. Bader, and B. Hillebrands, Appl. Phys. Lett. \textbf{101}, 042410 (2012).

\bibitem{klingler_design_2014} S. Klingler, P. Pirro, T. Br\"acher, B. Leven, B. Hillebrands, and A.V. Chumak, Appl. Phys. Lett. \textbf{105}, 152410 (2014).

\bibitem{gladii_spin_2016} O. Gladii, M. Collet, K. Garcia-Hernandez, C. Cheng, S. Xavier, P. Bortolotti, V. Cros, Y. Henry, J.-V. Kim, A. Anane, and M. Bailleul, Appl. Phys. Lett. \textbf{108}, 202407 (2016).

\bibitem{demokritov_brillouin_2001} S.O. Demokritov, B. Hillebrands, and A.N. Slavin, Phys. Rep. \textbf{348}, 441 (2001).

\bibitem{demidov_direct_2010} V.E. Demidov, S. Urazhdin, and S.O. Demokritov, Nature Mater. \textbf{9}, 984 (2010). 

\bibitem{Madami_2011} M. Madami, S. Bonetti, G. Consolo, S. Tacchi, G. Carlotti, G. Gubbiotti, F.B. Mancoff, M.A. Yar, and J. \AA kerman, Nature Nanotechnol. \textbf{6}, 635 (2011).

\bibitem{Chen_review} T. Chen, R.K. Dumas, A. Eklund, P.K. Muduli, A. Houshang, A.A. Awad, P. D\"urrenfeld, B.G. Malm, A. Rusu, and J. \AA kerman, Proc. IEEE \textbf{104}, 1919 (2016).

\bibitem{collet_generation_2016} M. Collet, X. de Milly, O. d'Allivy Kelly, V.V. Naletov, R. Bernard, P. Bortolotti, J. Ben Youssef, V.E. Demidov, S.O. Demokritov, J.L. Prieto, M. Mu\~{n}oz, V. Cros, A. Anane, G. de Loubens, and O. Klein, Nature Commun. \textbf{7}, 10377 (2016).

\bibitem{liu_spin_2007} Z. Liu, F. Giesen, X. Zhu, R.D. Sydora, and M.R. Freeman, Phys. Rev. Lett. \textbf{98}, 087201 (2007).

\bibitem{wu_fast_2006} M. Wu, B.A. Kalinikos, P. Krivosik, and C.E. Patton, J. Appl. Phys. \textbf{99}, 013901 (2006).

\bibitem{vlaminck_spin-wave_2010} V. Vlaminck and M. Bailleul, Phys. Rev. B \textbf{81}, 014425 (2010).

\bibitem{cherepov_electric-field-induced_2014} S. Cherepov, P.K. Amiri, J.G. Alzate, K. Wong, M. Lewis, P. Upadhyaya, J. Nath, M. Bao, A. Bur, T. Wu, G.P. Carman, A. Khitun, and K.L. Wang, Appl. Phys. Lett. \textbf{104}, 082403 (2014).

\bibitem{Barra} A. Barra, A. Mal, G.P. Carman, and A. Sepulveda, Appl. Phys. Lett. \textbf{110}, 072401 (2017).

\bibitem{Duflou} R. Duflou, F. Ciubotaru, A. Vaysset, M. Heyns, B. Sor\'ee, I.P. Radu, and C. Adelmann, Appl. Phys. Lett. \textbf{111}, 192411 (2017).

\bibitem{SOT_Review} P. Gambardella and I.M. Miron, Philos. Trans. Royal Soc. A \textbf{369}, 3175 (2011).

\bibitem{yu_switching_2014} G. Yu, P. Upadhyaya, Y. Fan, J.G. Alzate, W. Jiang, K.L. Wong, S. Takei, S.A. Bender, L.-T. Chang, Y. Jiang, M. Lang, J. Tang, Y. Wang, Y. Tserkovnyak, P.K. Amiri, and K.L. Wang, Nature Nanotechnol. \textbf{9}, 548 (2014).

\bibitem{fukami_spinorbit_2016} S. Fukami, T. Anekawa, C. Zhang, and H. Ohno, Nature Nanotechnol. \textbf{11}, 621 (2016).

\bibitem{fukami_sub-ns_2016} S. Fukami, T. Anekawa, A. Ohkawara, C. Zhang, and H. Ohno, Trans. IEEE Symp. VLSI Technol. (2016). DOI: \texttt{10.1109/VLSIT.2016.7573379}

\bibitem{baumgartner_time-_2017} M. Baumgartner, K. Garello, J. Mendil, C.O. Avci, E. Grimaldi, C. Murer, J. Feng, M. Gabureac, C. Stamm, Y. Acremann, S. Finizio, S. Wintz, J. Raabe, and P. Gambardella, Nature Nanotechnol. \textbf{12}, 980 (2017).

\bibitem{ciubotaru_all_2016} F. Ciubotaru, T. Devolder, M. Manfrini, C. Adelmann, and I.P. Radu, Appl. Phys. Lett. \textbf{109}, 012403 (2016).

\bibitem{bilzer_vector_2007} C. Bilzer, T. Devolder, P. Crozat, C. Chappert, S. Cardoso, and P.P. Freitas, J. Appl. Phys. \textbf{101}, 074505 (2007).

\bibitem{garello_symmetry_2013} K. Garello, I.M. Miron, C.O. Avci, F. Freimuth, G. Gaudin, O. Boulle, Y. Mokrousov, S. Bl\"ugel, S. Auffret, and P. Gambardella, Nature Nanotechnol. \textbf{8}, 587 (2013).

\bibitem{avci_interplay_2014} C.O. Avci, K. Garello, M. Gabureac, A. Ghosh, A. Fuhrer, S.F. Alvarado, and P. Gambardella, Phys. Rev. B \textbf{90}, 224427 (2014).

\bibitem{hayashi_quantitative_2014} M. Hayashi, J. Kim, M. Yamanouchi, and H. Ohno, Phys. Rev. B \textbf{89}, 144425 (2014).

\bibitem{liu_spin-torque_2012} L. Liu, C.-F. Pai, Y. Li, H.W. Tseng, D.C. Ralph, and R.A. Buhrman, Science \textbf{336}, 555 (2012).

\bibitem{Ta_relative_SOT} G. Allen, S. Manipatruni, D.E. Nikonov, M. Doczy, and I.A. Young, Phys. Rev. B \textbf{91}, 144412 (2015).

\bibitem{FL-SOT} X. Fan, H. Celik, J. Wu, C. Ni, K.-J. Lee, V.O. Lorenz, and J.Q. Xiao, Nature Commun. \textbf{5}, 3042 (2014).

\bibitem{OOMMF} The simulations were performed using the OOMMF open code: M. J. Donahue and D. G. Porter, NISTIR Report No. 6376 (1999, unpublished).

\bibitem{kalinikos_theory_1986} B.A. Kalinikos and A.N. Slavin, J. Phys. C \textbf{19}, 7013 (1986).

\bibitem{beta-W} C.-F. Pai, L. Liu, Y. Li, H.W. Tseng, D.C. Ralph, and R.A. Buhrman, Appl. Phys. Lett. \textbf{101}, 122404 (2012).

\bibitem{SOT_TI} A.R. Mellnik, J.S. Lee, A. Richardella, J.L. Grab, P.J. Mintun, M.H. Fischer, A. Vaezi, A. Manchon, E.-A. Kim, N. Samarth, and D.C. Ralph, Nature \textbf{511}, 449 (2014).

\bibitem{SOT_TI2} Y. Fan, P. Upadhyaya, X. Kou, M. Lang, S. Takei, Z. Wang, J. Tang, L. He, L.-T. Chang, M. Montazeri, G. Yu, W. Jiang, T. Nie, R.N. Schwartz, Y. Tserkovnyak, and K.L. Wang, Nature Mater. \textbf{13}, 699 (2014).

\bibitem{RFBOOK} T.H. Lee, \textit{Planar Microwave Engineering} (Cambridge Univ. Press, Cambridge, 2004).


\end{thebibliography}
\end{document}